\documentclass{PoS}

\def\lsim{\raise0.3ex\hbox{$<$\kern-0.75em\raise-1.1ex\hbox{$\sim$}}}
\def\gsim{\raise0.3ex\hbox{$>$\kern-0.75em\raise-1.1ex\hbox{$\sim$}}}

\title{On the universal critical behavior in 3-flavor QCD}

\ShortTitle{Critical behavior in 3-flavor QCD}

\author{\speaker{Dominik Smith}$^{\;a,c}$ and Christian Schmidt$^{\;b,c}$
\thanks{This work has been supported through the Helmholtz International 
Center for FAIR which is part of the Hessian initiative LOEWE.}\\
\llap{$^a$}Institut f\"ur theoretische Physik,
           J.W.Goethe Universit\"at Frankfurt, 
           D-60438 Frankfurt am Main, Germany\\
\llap{$^b$}Frankfurt Institute for Advanced Studies, 
           J.W.Goethe Universit\"at Frankfurt, 
           D-60438 Frankfurt am Main, Germany\\
\llap{$^c$}GSI Helmholtzzentrum f\"ur Schwerionenforschung,
           Planckstr.~1, D-64291 Darmstadt, Germany\\
E-mail: \email{smith@th.physik.uni-frankfurt.de}, 
\email{cschmidt@fias.uni-frankfurt.de}}

\abstract{We analyze the universal critical behavior at the chiral
critical point in QCD with three degenerate quark masses. We confirm
that this critical point lies in the universality class of the three
dimensional Ising model. The symmetry of the Ising model, which is
Z(2), is not directly realized in the QCD Hamiltonian. After making an
ansatz for the magnetization- and energy-like operators as linear
admixtures of the chiral condensate and the gluonic action, we
determine several non-universal mixing and normalization
constants. These parameters determine an unambiguous mapping of the
critical behavior in QCD to that of the 3d-Ising model. We verify its
validity by showing that the thus obtained orderparameter scales in
accordance with the magnetic equation of state of the 3d-Ising model.}

\FullConference{XXIX International Symposium on Lattice Field Theory \\
                 July 10-16 2011\\
                 Squaw Valley, Lake Tahoe, California}

\begin{document}
\section{Introduction}
A detailed understanding of the phase structure of strongly
interacting matter at non-zero temperature and density is one of the
major tasks for non-perturbative lattice QCD simulations. It has been
conjectured based on renormalization group arguments
\cite{Pisarski_Wilczek} and verified in numerical calculations
\cite{KLS,dFP_3f} that the QCD phase transition with three degenerate
quarks flavors is of first order for small masses.  This also holds
for two light quarks and one heavier strange quark, if the strange
quark mass is kept below a critical value. The line that separates the
first order region from the crossover region is a critical line on
which the nature of QCD phase transition is of second order
\cite{dFP_line}.  It is an open and interesting question how close
the critical line passes the physical point, where we know that the
QCD transition is a crossover \cite{Fodor_Nature}. In fact, recent lattice
calculations suggest that the first order region dramatically shrinks with
the approach of the continuum limit \cite{Endrodi,dFP_Nt6}.

In this work we restrict ourselves to three degenerate quark masses. 
Within this theory the critical point -- the endpoint of the line of first 
order transitions -- has been determined before, with standard staggered 
fermions on $N_\tau=4$ lattices \cite{KLS,dFP_3f} and on $N_\tau=6$ lattices 
\cite{dFP_Nt6}. Our analysis is also based on simulations with standard
staggered fermions (and the standard Wilson gauge action) on $N_\tau=4$ 
lattices. We repeat the determination 
of the critical endpoint with larger spatial lattice sizes. Moreover 
we focus on the determination of the correct scaling operators and 
variables that allow for the mapping of the QCD critical behavior onto
the corresponding universality class, which is here the 3-dimensional Ising
model. The construction of the operators and fields is done along the line
of \cite{KLS,mixing}.

\begin{figure}
\begin{center}
\resizebox{0.49\textwidth}{!}{%
\includegraphics{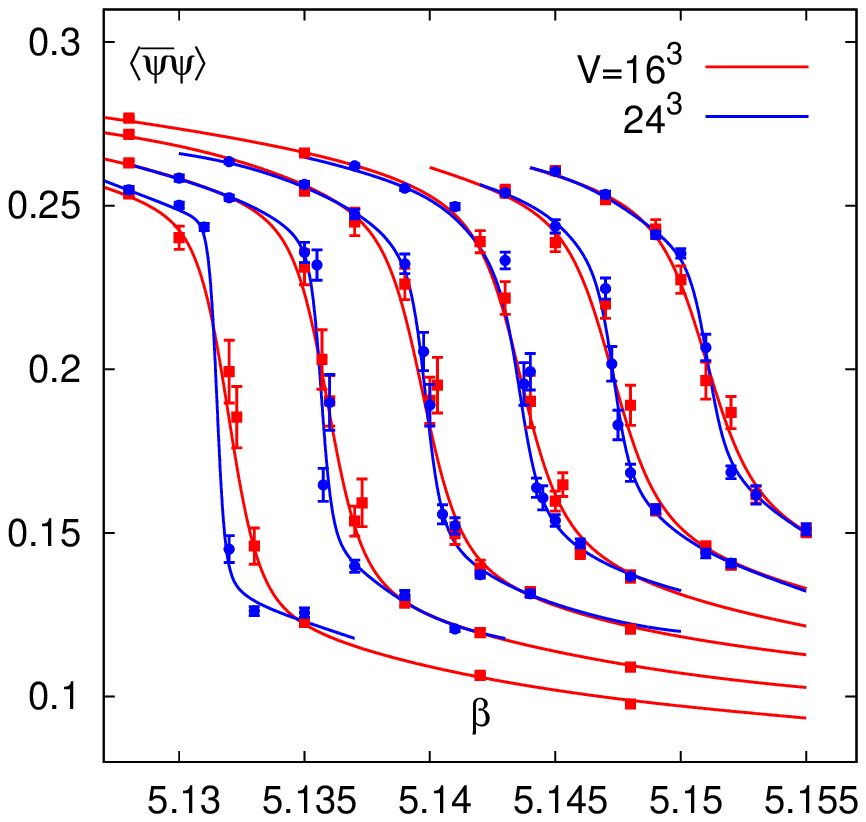}}
\resizebox{0.49\textwidth}{!}{%
\includegraphics{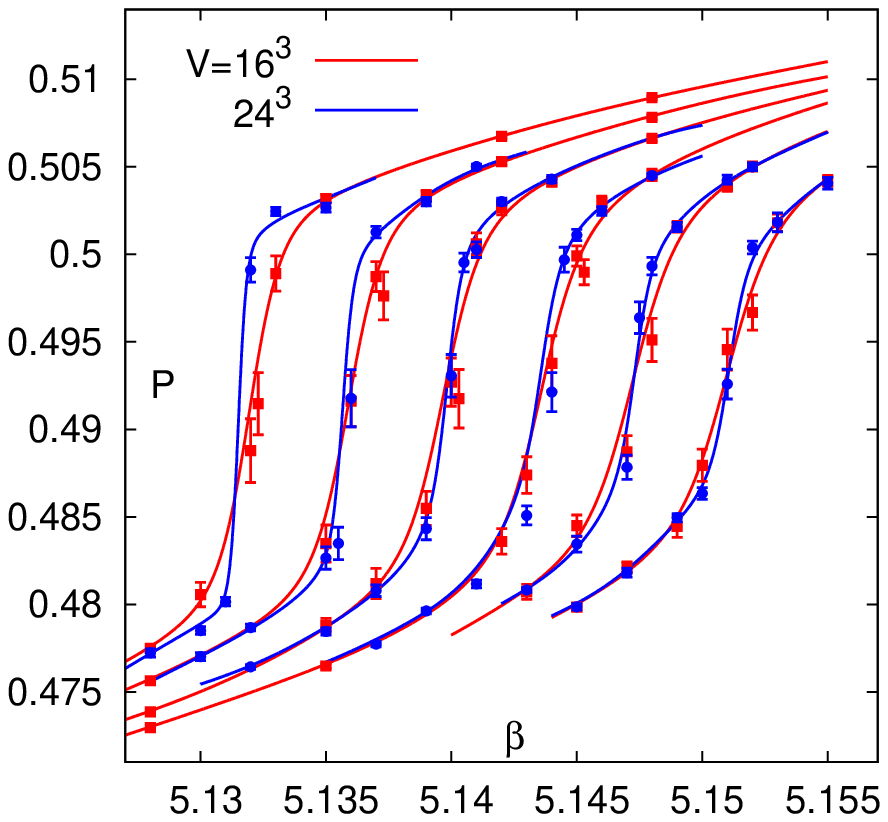}}
\end{center}
\caption{\label{fig:pbp_plaq}
The chiral condensate (left) and the plaquette expectation values for
different bare quark masses as a function of the coupling
$\beta=6/g^2$. The values of the bare quark masses are $am=0.024,
0.026, 0.028, 0.030, 0.032, 0.034$ (from left to right).}
\end{figure}

For our simulation we use an exact RHMC algorithm \cite{RHMC}, which
we have implemented to run entirely on GPUs. {\it I.e.}, all
calculations that involve spinors or gauge fields are done on the GPU,
whereas the CPU is only used to control the I/O and program
flow. The HMC trajectory is done in a mixed-precision manner: the
force and evolution calculations are done in single precision, which
is corrected by a Metropolis step in double precision. The expectation
values which we obtain for the chiral condensate and the plaquette are shown
in Fig.~\ref{fig:pbp_plaq}.
In oder to analyze the finite size scaling behavior we have performed 
simulations on spatial volumes of $16^3$, $24^3$ in some cases also $32^3$. 
For each coupling $\beta=6/g^2$ we have generated between 5000 and 10000 
configurations that are separated by 20 RHMC updates. 
For the bare quark masses, we picked values in the range of $0.024<am<0.034$.
From Fig.~\ref{fig:pbp_plaq} one can already see that the transition 
becomes stronger for smaller quark masses.

\section{Z(2) critical behavior}
The dynamics and universal critical behavior of the 3d-Ising model is
-- as in all generic O(N) models -- characterized by two relevant
scaling variables. Close to the critical point we thus can assume that
the Hamiltonian takes the effective form
\begin{equation}
H_{\rm eff}(t,h)=tE+hM\;,
\end{equation}
where $t$ and $h$ are the relevant scaling variables and $E,M$ the
corresponding scaling operators. In analogy to a classical spin system we 
will refer to the fields as reduced temperature and external field, and 
denote the operators as energy and magnetization. 

Under a renormalization group
transformation the singular part of the free energy scales as
\begin{equation}
f_s(t,h)=b^{-3}f_s(b^{y_t}t,b^{y_h}h)\;.
\label{eq:fs}
\end{equation}
The dimensionless scale factor $b$ is arbitrary. From here immediately
all known scaling and hyper-scaling relations follow. Especially the
standard finite size scaling behavior (choosing $b=LT=N_\sigma/N_\tau$)
of the susceptibilities is given by 
\begin{equation}
V^{-1}\left<(\delta E)^2\right>\sim L^{\alpha/\nu}\;, \quad
V^{-1}\left<(\delta M)^2\right>\sim L^{\gamma/\nu} \quad 
\mbox{with} \quad
\delta X= X -\left<X\right>\;.
\end{equation}
Here $L$ denotes the spatial extent of the system and $\alpha, \gamma,
\nu$ are the critical indices that characterize the universality
class. In order to map the QCD critical behavior to the Z(2) critical
behavior
we construct the scaling operators from the operators appearing in the 
QCD Lagrangian by employing the following linear ansatz \cite{KLS}
\begin{equation}
E=S_G+r\bar\psi\psi\;,\quad
M=\bar\psi\psi+sS_G\;.
\label{eq:ops}
\end{equation}
Here $S_G$ denotes the Wilson gauge action and $\bar\psi\psi$ the chiral 
condensate. Similarly, we linearize the scaling fields in the vicinity 
of the critical point $(\beta_c,m_c)$ by assuming
\begin{equation}
t=t_0((\beta-\beta_c)+A(m-m_c))\;, \quad
h=h_0((m-m_c)+B(\beta-\beta_c))\;.
\label{eq:dirs}
\end{equation}
The mixing parameters of the scaling operators and fields: $r,s,A,B$ as
well as the normalization constants $t_0,h_0$ have to be determined from 
the QCD simulations. Although there is no apparent reason, we assume 
in the following the directions $t$ and $h$ as orthogonal, which relates $A$
and $B$ by $A=tan(\phi)=-B$, where $\phi$ is the rotation angle between
the $\beta$ and the $t$ direction.

\section{Locating the critical point}
\begin{figure}
\begin{center}
\resizebox{0.49\textwidth}{!}{%
\includegraphics{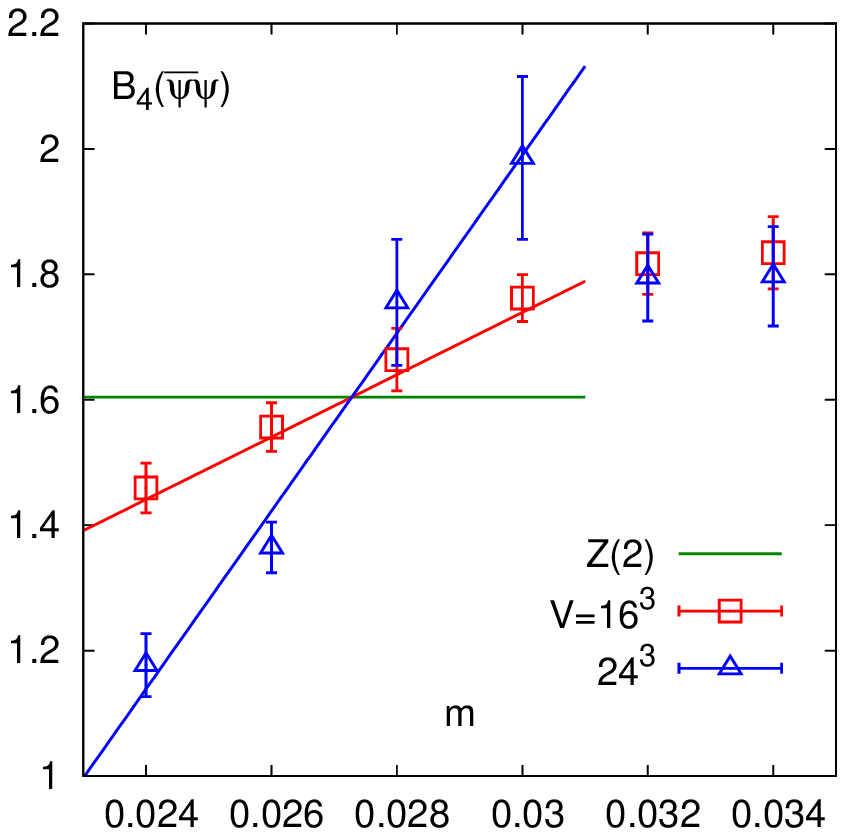}}
\resizebox{0.49\textwidth}{!}{%
\includegraphics{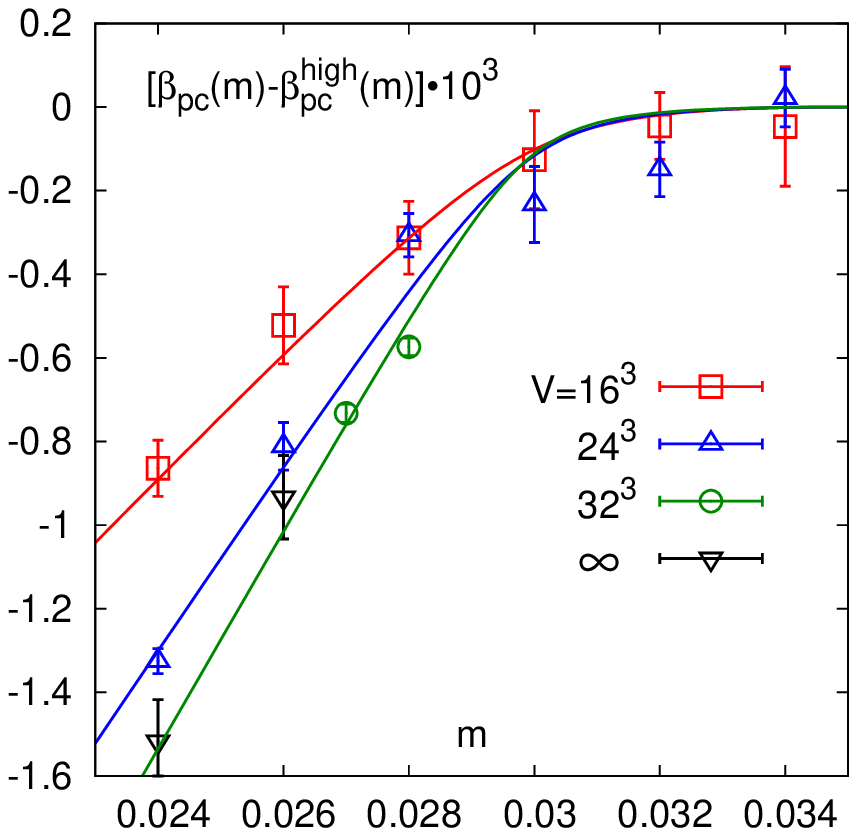}}
\end{center}
\caption{\label{fig:B4_betac}
The fourth order Binder cumulant of the chiral condensate (left) and the 
pseudo-critical coupling from the peak position of the disconnected chiral 
susceptibility (right).}
\end{figure}

The first step to establish the correct scaling operators and fields
in the 3-flavor QCD case is the location of the critical point. It has
has been shown \cite{KLS} that a good method to do so is by
analyzing Binder cumulants. The Binder cumulant of the order parameter
is defined by
\begin{equation}
B_4(M)=\left<(\delta M)^4\right>/\left<(\delta M)^2\right>^2\;.
\end{equation}
From the scaling behavior of the free energy (\ref{eq:fs}) it follows
immediately that the Binder cumulant is volume independent at the
critical point. Moreover, its value at the critical point is universal.
Furthermore, we can conclude form (\ref{eq:fs}) that close to the 
critical point
it is not mandatory to analyze the Binder cumulant of the correct order 
parameter. As long
as the operator of consideration has non-vanishing overlap with $M$, the 
strongest singularity corresponding to $L^{\gamma/\nu}$ will dominate and
the correct universal value of $B_4$ is obtained.
This means that in our analysis of Binder cumulants we can choose the 
mixing parameter $s$ arbitrarily\footnote{As long as we have not $s=1/r$, 
where the leading order singularity exactly cancels.}.
In Fig.~\ref{fig:B4_betac} (left) we plot the Binder cumulant of the 
chiral condensate ($s=0$) along the pseudo-critical line. 
{\it I.e.} for each mass we evaluate $B_4$ at the pseudo-critical
coupling $\beta_{pc}$(m) which we have determined from the peak position
of the chiral susceptibility. The two volumes $V=16^3$ and $24^3$
clearly suggest an intersection point close to the universal value of the
3d-Ising model, which is $B_4^{\rm Ising}=1.604$. We can thus confirm
that the critical point lies in the $Z(2)$ universality
class. Furthermore, we obtain for the intersection point of the $16^3$-lattice 
with the universal Ising value a mass value of $am_c=0.0268(6)$, which is in 
good agreement with the
previously obtained value from RHMC simulations
$am_c=0.0263(3)$\cite{dFP_line}. Combining the results of the $24^3$,
and $16^3$-lattices into a common fit that restricts the intersection
point of the two volumes to Ising universal value, we obtain a
critical mass of $am_c=0.0273(4)$. This fit is shown as straight lines
in Fig.~\ref{fig:B4_betac} (left). For both of the above described
fits we restrict the fitting range to $am \in [0.024,0.030]$. Larger
quark masses do not seem to fall into the scaling region, as the data
does not follow the approximate linear behavior for $am\ge 0.032$.

In Fig.~\ref{fig:B4_betac} (right) we show the pseudo critical
couplings for different volumes as function of mass. For better
visibility we have subtracted the linear function $\beta_{pc}^{\rm
high}(am)=1.8281\cdot am+5.0889$ and multiplied the differences by $10^3$.  
We see that for large masses ($am>am_c$) the pseudo critical couplings
seem to be volume independent within errors, whereas at sub-critical 
masses we observe a distinct scaling with the volume. Assuming that the
two masses $am=0.024$ and $0.026$ lie in the first order region we
perform to extrapolations of the pseudo critical couplings to the
thermodynamic limit, using
$\beta_{pc}(N_\sigma)=\beta_{c}(\infty)+cN_\sigma^{-3}$. 
Here $c$ is a constant that we fit.
We find that the infinite volume can, to a good approximation, already be 
described by the $32^3$-lattices.  
For $am<am_c$ we obtain a parameterization of the critical couplings in the
 thermodynamic limit as a function of the quark mass, as 
 $\beta_{c}(am)=2.12(3)\times (am-0.0268)5.1372(1)$. This is by
 definition a parameterization of the (negative) t-axis.
We can thus obtain a value of the mixing parameter
$B$ through the relation
\begin{equation}
B^{-1}=-\left. \frac{\partial \beta_{c}(am)}{\partial (am)}\right|_{am=am_c}\;,
\end{equation}
which arises due to the fact the in the Ising model the line of 
first order phase transitions defines the temperature direction. 
We estimate B=-0.47(1). Assuming that the scaling directions are orthogonal,
this also determines the parameter A.

\section{The mixing parameters $r$ and $s$}
So far we have determined the critical point as well as the proper scaling
directions. Before we can consider the magnetic equation of state we still 
have to clarify the correct order parameter $M$ through the determination
of the mixing parameter $s$. We fix $s$ by demanding that the linear 
combination $M=\bar\psi\psi+sS_G$ fulfills basic properties 
of an order parameter, namely that $M$ vanishes at the critical point and
stays zero for $t>0,h=0$. While the former condition is easy to fulfill and 
yields a mixing parameter $s=0.048(2)$, the latter condition is equivalent 
with the fact
that energy and magnetization operators are statistically independent:
\begin{equation}
\left<(\delta M) (\delta E)\right>=0\;.
\end{equation}
Using this relation together with the definition of the scaling operators
(\ref{eq:ops}) and directions (\ref{eq:dirs}), we obtain the following 
conditions for the mixing parameters:
\begin{equation}
r=-B\qquad \mbox{and}
\qquad s=\frac{\left<(\delta\bar\psi\psi)(\delta S_G)\right>
-B\left<(\delta\bar\psi\psi)^2\right>}{\left<(\delta S_G)^2\right>
-B\left<(\delta\bar\psi\psi)(\delta S_G)\right>}\;.
\label{eq:mix_s}
\end{equation}
As one can see, we require as input the already determined mixing parameter
$B$, as well as the susceptibilities and mixed susceptibilities of the 
QCD operators $\bar\psi\psi$ and $S_G$. 
As susceptibilities are usually much more difficult do determine, 
the error on the mixing parameter $s$ is much larger using this condition.
Moreover, the value for the mixing parameter we obtain from Eq.~(\ref{eq:mix_s})
is $s=-0.8(1)$, which is very different form the previously obtained value. We conclude
that with our ansatz for $M$ we cannot at the same time obtain a very flat behavior for 
$M(t)$ at $t>0$ and demand that $M(t)$ vanishes at $t=0$. This is most likely due 
to the finite volume effects that are still present in the $24^3$-lattices close to
the critical point. We note, that we can obtain a good fit to the magnetic equation
of state as described in the next section with both values of $s$. However, as long
as $M$ does not vanish at the critical point, one is forced to keep the leading order
contribution from the regular part of the free energy. This will contribute a constant
value to $M$ and lead to the desired behavior of the order parameter. 

\section{Magnetic equation of state}
\begin{figure}
\begin{center}
\resizebox{0.75\textwidth}{!}{%
\includegraphics{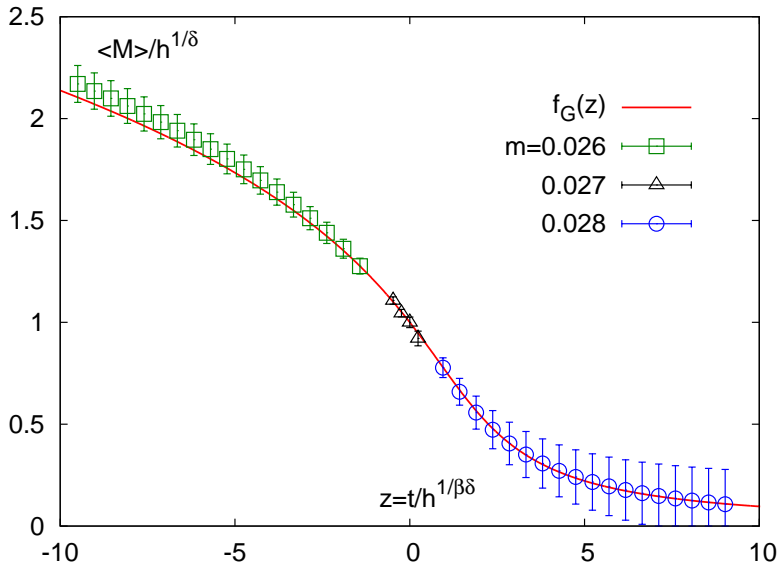}
}
\end{center}
\caption{\label{fig:fG}
The rescaled order parameter $M$ as function of the scaling 
variable $z$.}
\end{figure}

Close to the critical point, where all regular contributions to the 
free energy become negligible, it follows from (\ref{eq:fs}) that the 
singular part of the free energy can be expressed as a function of 
the single scaling parameter $z=t/h^{1/\delta}$ (by setting the scale
parameter $b$ to $b=h^{-{y_h}}$). For the order parameter (magnetization),
which is the derivative of the free energy with respect to the symmetry
breaking field $h$, one then obtains
\begin{equation}
M(t,h)=h^{1/\delta}f_G(z)\;.
\end{equation}
This relation is known as the magnetic equation of state. The scaling
function $f_G(z)$ is unique to the universality class. In the following we
use the parameterization as derived in Ref.~\cite{Zinn-Justin}. In order to obtain
the last two missing normalization constants $t_0$ and $h_0$, we fit our
data for the order parameter $M$ in the range of $0.026\le am \le 0.028$
to the magnetic equation of state. Using $s=0.048$, $am_c=0.0268$ and $\beta_c=5.1372$ 
we obtain $t_0=0.0210(9)$ and $h_0=0.0005(1)$. In Fig.~\ref{fig:fG} we plot the rescaled
order parameter $M/h^{1/\delta}$ as a function of the scaling parameter $z$.
We find good scaling behavior within the range of 
$-10\le z \le 10$. 

\section{Outlook}
In our future work we will address the question whether the good
scaling behavior of our constructed order parameter can also be observed 
in corresponding susceptibilities. Of special interest will be not only the
chiral susceptibility but also the mixed susceptibility that contains
the response to the variation of a small chemical potential
($\partial^2\ln Z/\partial \mu \partial h\equiv \partial M/\partial
\mu $). From this susceptibility we will be able to determine the
curvature of the critical surface that is defined by the extension of
the critical line in the quark mass plane towards non-zero chemical
potentials.


\begin{thebibliography}{99}
\bibitem{Pisarski_Wilczek}
R.~Pisarski and F.~Wilczek, \emph{Phys. Rev. D} {\bf 29} (1984) 338.
\bibitem{KLS}
  F.~Karsch, E.~Laermann, C.~Schmidt,
  Phys.\ Lett.\  {\bf B520 } (2001) 41
  [hep-lat/0107020].
\bibitem{dFP_3f}
  P.~de Forcrand, O.~Philipsen,
  Nucl.\ Phys.\  {\bf B673 } (2003)  170
  [hep-lat/0307020].
\bibitem{dFP_line}
  P.~de Forcrand, O.~Philipsen,
  JHEP {\bf 0701 } (2007) 077
  [hep-lat/0607017].
\bibitem{Fodor_Nature}
  Y.~Aoki, G.~Endrodi, Z.~Fodor, S.~D.~Katz, K.~K.~Szabo,
  Nature {\bf 443 } (2006) 675
  [hep-lat/0611014].
\bibitem{Endrodi}
  G.~Endrodi, Z.~Fodor, S.~D.~Katz, K.~K.~Szabo,
  PoS {\bf LAT2007} (2007)  228 
  [arXiv:0710.4197 [hep-lat]].
\bibitem{dFP_Nt6}
  P.~de Forcrand, S.~Kim, O.~Philipsen,
  PoS {\bf LAT2007 } (2007)  178
  [arXiv:0711.0262 [hep-lat]].
\bibitem{mixing}
J.~J.~Rehr and N.~D.~Mermin, Phys.\ Rev.\ {\bf A8} (1973) 472;
N.~B.~Wilding, J. Phys.: Condens. Matter {\bf 9} (1997) 585.
\bibitem{RHMC}
  A.~D.~Kennedy, I.~Horvath, S.~Sint,
  Nucl.\ Phys.\ Proc.\ Suppl.\  {\bf 73 } (1999)  834-836.
  [hep-lat/9809092].
\bibitem{Zinn-Justin}
  J.~Zinn-Justin, Phys.\ Rept.\ 344 (2001) 159 [hep-th/0002136].
\end{thebibliography}
\end{document}